\documentclass[sigconf, screen, nonacm]{acmart}
\renewcommand\footnotetextcopyrightpermission[1]{} 
\usepackage{amsmath,amssymb,amsfonts}
\usepackage{amsthm}
\usepackage{mathrsfs}

\usepackage{graphicx}
\usepackage{tikz}
\usetikzlibrary{quantikz2}
\usetikzlibrary{trees}
\usepackage{pgfplots}
\pgfplotsset{compat=1.18}

\usepackage{multirow}
\usepackage{array}
\usepackage{booktabs}
\usepackage{float}
\usepackage{subcaption}

\usepackage{algorithm}
\usepackage{algorithmicx}
\usepackage{algpseudocode}
\usepackage{listings}

\usepackage[title]{appendix}
\usepackage{xcolor}
\usepackage{textcomp}
\usepackage{manyfoot}
\usepackage{comment}
\usepackage{natbib}

\usepackage{url}
\usepackage{hyperref}

\setcopyright{none}
\begin{document}

\title{Quantum State Preparation with the QNN-based SRBB Algorithm}


\author{Marco Mordacci}
\authornote{Both authors contributed equally to this research.}
\author{Giacomo Belli}
\authornotemark[1]
\author{Michele Amoretti}

\affiliation{%
  \institution{Quantum Software Laboratory\\University of Parma}
  \city{Parma}
  \country{Italy}}
\email{{giacomo.belli,marco.mordacci1,michele.amoretti}@unipr.it}


\begin{abstract}In this work, a novel algorithm structured on Lie algebras for the approximate quantum state preparation problem is proposed, addressing a challenge of fundamental importance in many areas of quantum computing. The algorithm uses a variational quantum circuit designed on the Standard Recursive Block Basis (SRBB), a hierarchical construction for the matrix algebra of the $SU(2^n)$ group, which is capable of linking the variational parameters with the topology of the Lie group. Compared to the full algebra, using only diagonal components reduces the number of CNOTs by an exponential factor, as well as the circuit depth, in full agreement with the relaxation principle inherent to the approximation methodology of minimizing resources while achieving high accuracy. The desired quantum state is then approximated by a novel quantum neural network, which is designed based on the diagonal SRBB sub-algebra. This approach provides a new scheme for approximate quantum state preparation in a variational framework and a specific use case for the SRBB hierarchy. 

The performance of the algorithm is assessed with different loss functions, such as fidelity, trace distance, and Frobenius norm, in relation to two optimizers: Adam and Nelder-Mead. The results highlight the potential of SRBB in close connection with the geometry of unitary groups, achieving high accuracy of up to 4 qubits in simulation, but also its current limitations with an increasing number of qubits. Additionally, the approximate SRBB-based QSP algorithm has been tested on real quantum devices to assess its performance with a small number of qubits.
\end{abstract}



\maketitle

\section{Introduction}\label{sec1}
Quantum state preparation (QSP) is a very important step in many quantum algorithms~\cite{Iaconis2024}, including various algorithms for quantum machine learning (QML)~\cite{rath2024quantum}, quantum algorithms to solve systems of linear equations~\cite{bravo2023variational}, and Hamiltonian simulation~\cite{berry2018improved}.
The goal of quantum state preparation is to construct a unitary operator $U$ such that $|\psi\rangle=U|0\rangle^{\otimes n}$, where $|0\rangle^{\otimes n}$ is the $n$-qubit initial state and $|\psi\rangle=\sum_{i=0}^{2^n-1}c_i|i\rangle$ represents the desired quantum state in the computational basis, satisfying the complex normalization condition $\sum_{i=0}^{2^n-1}|c_i|^2=1$. Its approximate version aims to prepare a state $|\phi\rangle$ which is $\epsilon$-close to $|\psi\rangle$ with respect to some metric. When defining quantum state preparation algorithms, the use of ancillary qubits is another aspect that must be taken into consideration. Below, to the best of our knowledge, the main results in the QSP research stream are summarized, dividing the contributions between techniques without and with ancillary qubits, QML methods, and research related to Gate Synthesis.

The study of QSP started in its exact version, without ancillary qubits, using the method proposed by Nielsen and Chuang~\cite{nielsen2000quantum} (depth upper bound of $\mathcal{O}(2^n)$). In the same period, while Grover and Rudolph~\cite{grover2002creating} proposed an algorithm for QSP in the special case of integrable probability density functions with a depth upper bound of $\mathcal{O}(n2^n)$, Long and Sun~\cite{long2001efficient} presented a scheme for arbitrary superpositions of the basis states, which required $\mathcal{O}(n^22^n)$ 1-qubit and controlled gates. In 2005, M\"ott\"onen et al.~\cite{mottonen2005transformation} proposed a QSP algorithm exploiting Uniformly Controlled Rotations (UCRs)~\cite{mottonen2004quantum}, with a depth of $\mathcal{O}(2^n)$ and $2^{n+2}-4n-4$ CNOT gates. Immediately after, Bergholm et al.~\cite{bergholm2005quantum} formalized the Uniformly Controlled Gate (UCG) class and provided an upper bound of $2^{n+1}-2n-2$ for the number of CNOTs, with depth also of order $\mathcal{O}(2^n)$, while Soklakov and Schack~\cite{soklakov2006efficient} obtained a very efficient algorithm for preparing sequences of states with suitably bounded amplitudes, following Grover's research line. In 2011, based on the universal gate decomposition proposed by M\"ott\"onen and Vartiainen~\cite{mottonen12006decompositions,vartiainen2004efficient}, Plesch and Brukner~\cite{plesch2011quantum} focused on minimizing the CNOT count, achieving $\frac{23}{24}2^n-2^{\frac{n}{2}+1}+\frac{5}{3}$ for even $n$ and $\frac{115}{96}2^n$ for odd $n$. Subsequently, QSP algorithms were formally incorporated into the systematic synthesis of isometries, thanks to the work of Iten et al.~\cite{iten2016quantum}, who proposed three decompositions close to the theoretical lower bound. Their work consolidated the already known link between Gate Synthesis and QSP based on the search for optimal decompositions and the modularity of the UCG class~\cite{shende2005synthesis}.

In 2016, by means of a synthesis approach to the QSP problem, Neimann et al.~\cite{niemann2016logic} derived a methodology to automatically synthesize a circuit capable of generating any desired $n$-qubit quantum state with $n$ Hadamard gates, $2^{n+1}+3(n-1)-4$ CNOT gates, and $2^{n+1}-2$ elementary rotations. Subsequently, Zhao et al.~\cite{zhao2019state} proposed both deterministic and probabilistic state preparation based on quantum phase estimation, linking QSP to the decomposition of diagonal unitary operators~\cite{zhang2024depth}. In 2020, Mozafari et al.~\cite{mozafari2020automatic,mozafari2022efficient} opened a research line on uniform QSP, where scalability is ensured by the binary decision diagram and the decompositions made possible by UCRs. This line has continued with the recent works of Tanaka et al.~\cite{tanaka2024quantum} and Hong et al.~\cite{hong2025advancing}. Recently, Ramacciotti et al.~\cite{ramacciotti2024simple} slightly modified the Grover-Rudolph algorithm for sparse quantum states, showing that the gate complexity is linear in the number of nonzero amplitudes and also in the number of qubits. In the same year, Mao et al.~\cite{mao2024toward} defined an algorithm with $\mathcal{O}(\frac{ns}{\log_2n+n})$ total gates for $s$-sparse arbitrary states.

The use of ancillae in the history of QSP began with Zhang et al.~\cite{zhang2021low}, whose algorithm can generate the target state in $\mathcal{O}(n^2)$ depth with $\mathcal{O}(2^{2n})$ ancillary qubits, but only with a limited success probability. In the same year, Rosenthal~\cite{rosenthal2021query} tackled the topic of space-time trade-off by obtaining an algorithm with depth $\mathcal{O}(n)$ using $\mathcal{O}(n2^n)$ ancillary qubits. In 2022, Zhang et al.~\cite{zhang2022quantum} presented another QSP circuit of depth $\mathcal{O}(n)$ using $\Theta(2^n)$ ancillary qubits, but the \emph{asymptotically} optimal space-time trade-off bounds were found by Sun et al.~\cite{sun2023asymptotically} in their algorithm with depth $\tilde{\mathcal{O}}(\frac{2^n}{m+n}+n)$ and width $\mathcal{O}(2^n)$. The line of research culminated in the work of Yuan et al.~\cite{yuan2023optimal}. Regarding the analysis of state preparation for sparse states with ancillae, a recent result was obtained by Li et al.~\cite{li2025nearly} through a systematic investigation of the circuit width. Furthermore, the work of Gui et al.~\cite{gui2024spacetime} opens an interesting perspective thanks to the introduction of another metric called \emph{spacetime allocation}, which aims to quantify how long each qubit remains active and the availability of free ancillae.

Narrowing the context, the following contributions should be highlighted within the field of Quantum Machine Learning, which seeks to utilize quantum computing to improve machine learning results in terms of performance and training efficiency~\cite{cerezo2021variational}. A Quantum Neural Network (QNN) is a hybrid quantum-classical approach comprising a Variational Quantum Circuit (VQC)~\cite{benedetti2019parameterized} and an optimization process performed using classical methods such as classical gradient descent. The VQC, which is the quantum counterpart of the classical neural network, is composed of a set of rotations whose angles are trained by the optimization process. In 2001, Kaye et al.~\cite{kaye2001quantum} proposed a quantum network that is very efficient for generating pure symmetric states, and in 2021, Caro et al.~\cite{caro2021encoding} derived generalization bounds for parameterized quantum circuits that depend explicitly on the strategy used for the encoding. Recently, Hai et al.~\cite{hai2023variational} introduced a variational approach to QSP for entangled states, where the parametric ansatz has a hypergraph-based structure. The variational methodology was further developed by Wang et al.~\cite{wang2023robuststate} with noise-aware training that exploits back-propagation and incorporates real quantum noise into gradient calculations. The strategy of approximating the desired state by means of a parametric circuit is also used by Zhao et al.~\cite{zhao2024superencoder}, while Gosset et al.~\cite{gosset2024quantum} have recently published a study on the optimal number of $T$ gates to approximate an arbitrary quantum state, improving previous work~\cite{low2024trading}. Furthermore, Mitsuda et al.~\cite{mitsuda2024approximate} generalized the approximate amplitude encoding (AAE) method to complex-data vectors.

Other interesting works worthy of mention are the following. In 2019, Sanders et al.~\cite{sanders2019black} presented a new algorithm for the class of black-box QSP that avoids traditional Grover's arithmetic, which is a key contributor to complexity. The work of Romero et al.~\cite{romero2017quantum}, Bondarenko et al.~\cite{bondarenko2020quantum}, and Zhang et al.~\cite{zhang2021generic} on quantum autoencoders (qae) poses interesting ideas for QSP, since once the qae has been trained, the decompression unitary can approximate the originally compressed states. Finally, Araujo et al.~\cite{araujo2021divide} proposed a combination of M\"ott\"onen's algorithm and a divide-and-conquer strategy to obtain an exponential speedup in the encoding process.

As stated in its own definition, the problem of preparing the quantum state is closely connected with the topic of Unitary Synthesis (or Gate Synthesis), exact or approximate~\cite{shende2004minimal,shende2005synthesis,barenco1995elementary,kliuchnikov2015framework,vatan2004optimal,kliuchnikov2012fast,jiang2020optimal}. Recently, an approach to approximate synthesis based
on Lie algebras~\cite{kirillov2008introduction} and classical optimization techniques was proposed~\cite{sarkar2023scalable}, building on the previous work of Madden and Simonetto~\cite{madden2022best}. The latter introduces a Hermitian unitary basis for the $\mathbb{C}^{2^n\times2^n}$ algebra, called the Standard Recursive Block Basis (SRBB), to obtain a scalable and recursive representation for any unitary operator. Subsequently, the scalable structure of the SRBB-based unitary decomposition was used to design the VQC of a quantum neural network for approximate synthesis tasks, optimized in terms of CNOTs, and tested both in simulation and with a real quantum computer~\cite{belli2024scalable,10821064}.

This paper provides the following new contributions: 1) a novel approximate QSP algorithm is proposed as a specific use case of the SRBB hierarchy; 2) the traditional ladder structure of QSP circuits is implemented with only the CNOT-optimized diagonal SRBB sub-algebra to reduce its depth, approaching the asymptotically optimal bounds of Sun et al.~\cite{sun2023asymptotically}; 3) the algorithm's performance is assessed with random states, sparse states, and states with a uniform probability distribution, up to 8 qubits; 4) the QNN is also tested with real quantum devices (provided by IQM~\cite{IQM} and AWS Braket) to assess its performance in a real environment with random states and specific states, such as Bell and GHZ.

This is the extended version, enriched with proofs and results on real devices, of the conference paper titled
``\emph{SRBB-Based Quantum State Preparation}'' that was presented at Computing Frontiers 2025 (CF'25).

The remainder of the paper is organized as follows. In Section~\ref{lab:theoretical_framewowk}, the \emph{Standard Recursive Block Basis} is formally introduced as the basis of the Lie algebra for the group $U(N)$, focusing on its diagonal sub-algebra represented by the $Z(\Theta_Z)$-factor and on the topological bundle structure of classical unitary groups. This theoretical framework, which describes the unitary group under algebraic, geometric, and topological aspects, is then exploited to address the state preparation problem, analyzing several possible implementation strategies related to the QSP natural structure based on Uniformly Controlled Gates (UCGs). In the last part of the section, it is shown, also through examples, how to build a scalable QSP circuit using only the SRBB $Z$-factor, $S$, and $H$ gates. Then, formulas for depth and gate count are provided. In Section~\ref{lab:results}, the implementation of the corresponding QNN is explained in detail. In addition, the results obtained with simulations and with real quantum devices are presented with a related description. Finally, Section~\ref{lab:conclusion} concludes the paper with a summary of the main results and a discussion of future work.

\section{Theoretical Framework}
\label{lab:theoretical_framewowk}
The \emph{Standard Recursive Block Basis} (SRBB)~\cite{sarkar2023scalable} is the hierarchical and recursive construction that defines for any $n\in\mathbb{N}_+$ the Hermitian unitary basis of the $\mathbb{C}^{2^n\times2^n}$ matrix algebra\footnote{The complex matrix algebra, often denoted by $M_N(\mathbb{C})$, is the vector space of all $N\times N$ matrices with complex entries, endowed with the usual operations of matrix addition and scalar multiplication over $\mathbb{C}$. A basis of this algebra is any set of $N^2$ linearly independent matrices that spans $M_N(\mathbb{C})$~\cite{horn2012matrix}.}. Originally introduced to obtain a scalable, parameterized representation of unitary matrices\footnote{A complex square matrix $U\in M_N(\mathbb{C})$ is unitary if $U^\dag U=UU^\dag=\mathbb{I}_N$, where the symbol $\dagger$ indicates the Hermitian conjugation and $\mathbb{I}_N$ is the identity matrix of order $N$. From a geometric perspective, $U$ is a complex isometry that preserves the standard Hermitian inner product $\langle v,w\rangle=v^\dag w$, for any $v,w\in\mathbb{C}^N$. The set of all unitary matrices of order $N$, denoted by $U(N)$, is a closed and compact subset of $M_N(\mathbb{C})$~\cite{reed1980methods}.}, it can be thought of as a generalization of the Pauli basis for complex matrices of higher orders. The set of unitary matrices is denoted by $U(2^n)=\{U\in M_{2^n}(\mathbb{C})\;|\;U^\dag U=\mathbb{I}_{2^n}\}$ and satisfies the group structure with respect to the matrix product. From a topological point of view, $U(2^n)$ is a compact and connected\footnote{A Lie group is called compact if it is bounded and closed in its ambient Euclidean space $M_N(\mathbb{C})\cong\mathbb{R}^{2N^2}$, hence topologically compact, and connected if any two points of the group can be joined by a continuous path lying entirely within the group~\cite{hall2013lie}.} Lie group of dimension $2^{2n}$, i.e., a group that is also a differentiable space (smooth manifold) where the multiplication and inversion operations are smooth~\cite{hall2013lie}. As a basis for the algebra of unitary matrices of order $2^n$, SRBB is the basis of a Lie algebra. A Lie group can be thought of as a ``curved'' space with a \emph{local} structure similar to a vector space: the Lie algebra is precisely the (linear) approximation of the group around the identity. A little more formally, the Lie algebra $\mathfrak{u}(2^n)$ associated with $U(2^n)$ is the tangent space\footnote{The Lie algebra $\mathfrak{u}(N)$ is a real vector space of dimension $N^2$, closed under the commutator $[X,Y]=XY-YX\in\mathfrak{u}(N)$. The notation $\mathfrak{u}(N)\simeq\mathbb{R}^{N^2}$ is sometimes used to indicate local or topological equivalence.} to the group identity, from which the convention on anti-Hermitian generators derives, $\mathfrak{u}(2^n)=\{X\in M_N(\mathbb{C})\;|\;X^\dag=-X\}$~\cite{hall2013lie}. This convention guarantees that the exponential $e^X\in U(2^n)$ has a real argument $X\in\mathfrak{u}(2^n)$: $e^X$ is unitary if $X$ is anti-Hermitian. While the group $U(2^n)$ describes finite transformations, the algebra $\mathfrak{u}(2^n)$ describes infinitesimal transformations close to the identity. The exponential map $\mbox{exp}:\mathfrak{u}(2^n)\rightarrow U(2^n)$ connects the two instances, transforming an infinitesimal generator into a finite transformation. In a similar way, the special subgroup $SU(2^n)$ is linked to the Lie algebra $\mathfrak{su}(2^n)=\{X\in M_N(\mathbb{C})\;|\;X^\dag=-X,\;\mbox{Tr}(X)=0\}$, given that $\det(e^X)=e^{\mbox{Tr}(X)}$.

Usually, in physical contexts, it is preferable to adopt a different convention and choose Hermitian generators to remain consistent with quantum observables~\cite{peskin2018introduction}. A fundamental property of $M_{2^n}(\mathbb{C})$, or rather of its Hermitian subclass, is that a Hermitian matrix $H^\dag=H$ can always be uniquely decomposed into a traceless component and another one proportional to the identity\footnote{In the decomposition, the generator proportional to the identity, $Y_{det}=\frac{\mathbb{I}_{2^n}}{2^n}$, is chosen according to the determinant convention so that to have $\mbox{Tr}(Y_{det})=1$ and $\det(e^{i\alpha Y_{det}})=e^{i\alpha}$ with $\alpha=\mbox{Tr}(H)$.}:
\begin{equation}
H=\left(H-\frac{\mbox{Tr}(H)}{2^n}\mathbb{I}_{2^n}\right)+\frac{\mbox{Tr}(H)}{2^n}\mathbb{I}_{2^n}
\end{equation}
While the first term with zero trace is the component in $\mathfrak{su}(2^n)_{phys}$, the second one is the component in the subspace $\mbox{span}_{\mathbb{R}}(\mathbb{I}_{2^n})$. Formally, the Hermitian subclass of $M_{2^n}(\mathbb{C})$ is defined as $\mbox{Herm}(2^n)=\mbox{span}_{\mathbb{R}}\{\mathbb{I}_{2^n}\}\oplus\mbox{span}_{\mathbb{R}}\{T_a\;,\;a=1,...,2^{2n}-1\;|\;T_a^\dag=T_a\;,\;\mbox{Tr}(T_a)=0\}$, which in group geometry translates into the \emph{local decomposition}\footnote{With the adjective ``local'', we refer to the smooth manifold or Lie algebra in the sense of differentiable structure with infinitesimal generators.} of the Lie group $U(2^n)$~\cite{nakahara2018geometry}:
\begin{equation}\label{eqn:decomp_algebra}
    \mathfrak{u}(2^n)_{phys}=\mathfrak{u}(1)\oplus\mathfrak{su}(2^n)_{phys}
\end{equation}
where the subscript ``$phys$'' indicates the convention in the physical field. The relationship between the two conventions is defined by the map between the generators, $X=iT$, or equivalently by $\mathfrak{su}(2^n)_{math}=i\cdot\mathfrak{su}(2^n)_{phys}$, such that $e^{iT}$ is unitary when $T$ is Hermitian. Accordingly, any unitary operator can be expressed as a product of exponentials of SRBB elements, ordered according to a specific algebraic criterion for implementation purposes~\cite{belli2024scalable}.

The decomposition of the $\mathfrak{u}(2^n)$ algebra in Eq.~(\ref{eqn:decomp_algebra}) induces the \emph{local group structure} via direct product~\cite{nakahara2018geometry}:
\begin{equation}\label{eqn:local_structure}
    U(N)\simeq U(1)\times SU(N) 
\end{equation}
where the symbol ``$\simeq$'' indicates that the relation holds only locally, that is, at the level of Lie algebras, and emphasizes a richer global group structure (or topology) defined below.

By virtue of the infinitesimal generators of the physical convention, any element $U\in U(2^n)$ can be written as a \emph{local parameterization}\footnote{In the local parameterization, the exponential map $\mathfrak{u}(N)\rightarrow U(N)$ is used in a neighborhood of the identity, choosing coordinates $(\alpha,\theta^a)$ on the Lie algebra. The map is unambiguous for small $|\alpha|$ as a result of the local choice, but globally, $\alpha$ is defined $\mbox{mod }2\pi$.} close to the identity,
\begin{equation}\label{eqn:local_chart}
U\simeq e^{i\alpha Y}e^{iH_0}\in U(1)\times SU(N)
\end{equation}
where $Y\equiv Y_{det}=\frac{\mathbb{I}_{2^n}}{2^n}$ and $H_0=\sum_a\theta^aT_a\in\mathfrak{su}(2^n)_{phys}$, in such a way that $\det(U)=\det(e^{i\alpha Y})\det(e^{iH_0})=e^{i\alpha}$, for $\alpha=\mbox{arg}\det(U)$. In other words, the determinant map encodes the global phase of the $U(1)$ group internal to $U(2^n)$. In differential geometry, this corresponds to the natural projective map $\pi:U(2^n)\rightarrow U(1)$, with $\pi(U)=\det(U)$ and $\mbox{ker}(\det)=SU(2^n)$. However, the decomposition expressed by Eq.~(\ref{eqn:local_chart}) is not unique: it is possible to change $\alpha$ into multiples of $2\pi$, obtaining new coordinates $(\alpha,\theta^a_{new})$ that return the same $U\in U(2^n)$, but with $H_0$ modified by a so-called \emph{central twist} (or phase torsion)~\cite{helgason1979differential}:
\begin{equation}
U\simeq e^{i(\alpha+2\pi k)Y}e^{iH_0}=e^{i\alpha Y}e^{i\left(\frac{2\pi k}{2^n}\mathbb{I}_{2^n}+H_0\right)}\quad\mbox{for all } k\in\mathbb{Z}
\end{equation}
So, although $\alpha$ is defined globally $\mbox{mod }2\pi$, when we restrict to $SU(2^n)$, it is defined $\mbox{mod }\frac{2\pi}{2^n}$. This is an indicator of the presence of a topological twist in 1-1 correspondence with the roots of $\det(U)$. From the point of view of the determinant, i.e., of the natural projective map, the parameter $\alpha$ is well defined on the $U(1)$ circle; in fact, $\det(U)=e^{i\alpha},\;\forall\alpha\in\mathbb{R}\mbox{ mod }2\pi$ is even far from the identity. If one wants to rewrite an element of $U(2^n)$ as a global phase multiplied by an element of $SU(2^n)$, then one must consider $U=e^{i\phi}S$ with $S\in SU(2^n)$ and $\phi\in\mathbb{R}$ such that
\begin{equation}
\det(U)=\det(e^{i\phi}S)=e^{2^n(i\phi)}\det(S)=e^{i\phi2^n}
\end{equation}
which implies $\phi=\frac{\alpha+2\pi k}{2^n}$ for $k=0,1,...,2^n-1$. At the global level, the discrete degeneracy of the roots of $\det(U)$ appears clearly and corresponds to the phase torsions within the $SU$ subspace\footnote{The n-th roots of $\det(U)$ were not introduced by analytical artifice, but derive from the multiplicative structure of the group and are the topological manifestation of the fact that $U(N)$ is not, globally, a direct product. It is precisely the discrete torsions within $SU(N)$ that introduces the equivalence modulo $\frac{2\pi}{N}$.}. In the language of generators, Eq.~(\ref{eqn:local_chart}) is equivalent to choosing a neighborhood of the identity where $k=0$. Therefore, a unitary matrix can be written as
\begin{equation}
U=e^{i(\alpha+2\pi k)/2^n}S_k\;,\quad S_k\in SU(2^n)
\end{equation}
For any $U\in U(2^n)$, there exists a family of $SU(2^n)$ matrices in one-to-one correspondence with the roots of its determinant in the formula
$$
S_k=e^{-i(\alpha+2\pi k)/2^n}U=e^{-i2\pi k/2^n}e^{-i\alpha}U=\left(e^{-i2\pi k/2^n}\mathbb{I}_{2^n}\right)S_0
$$
where each $S_k$ differs from the others by a discrete phase $e^{-i\frac{2\pi k}{2^n}}$, that is, by an element of the so-called \emph{discrete center} $\mathbb{Z}_{2^n}$. In general, the center $Z(G)$ of a group $G$ is the set of elements that commute with the entire group: $Z(G)=\{g\in G\;|\;gh=hg\;,\;\forall h\in G\}$. If $G=SU(2^n)$, it is shown that $Z=\{e^{i2\pi k/2^n}\mathbb{I}_{2^n}\;|\;k=0,1,...,2^n-1\}$, since the discrete global phases commute with everything, being proportional to the identity, and their determinant is 1~\cite{helgason1979differential}. In other words, $SU(2^n)$ has a finite subgroup of discrete global phases isomorphic to $\mathbb{Z}_{2^n}$. From a physical point of view, it is the set of indistinguishable transformations, that is, those that do not produce an effect on observables. In fact, as just shown, writing $U=e^{i\alpha_k}S_k\in U(2^n)$, where $\alpha_k$ is one of the roots of $\det(U)$, hides a redundancy: if you multiply $S$ by an element of the center $z\in Z(SU)$, and simultaneously multiply $e^{i\alpha_k}$ by $z^{-1}$, you get the same unitary matrix. Formally, it is like having an equivalence class for each $U=e^{i\alpha_k}S_k$, where the set of equivalence classes is defined by: $[(S,e^{i\alpha_k})]=\{(zS,e^{i\alpha_k}z^{-1})\;|\;z\in Z(SU)\}$. This redundancy shows that the global structure of $U(2^n)$ cannot be described by a direct product, but rather by a \emph{quotient product} over $\mathbb{Z}_{2^n}$ (or central product over $\mathbb{Z}_{2^n}$), since it identifies pairs that differ by an element of the discrete center $\mathbb{Z}_{2^n}\subset SU(2^n)$:
\begin{equation}
U(2^n)\cong[SU(2^n)\times U(1)]/\mathbb{Z}_{2^n}
\end{equation}
where the symbol ``$\cong$'' stands for global isomorphism~\cite{nakahara2018geometry}. The twists, or phase torsions, of $SU$ related to the multiplicity of the roots of the determinant are a topological manifestation of the equivalence classes, thanks to the structural role played by the discrete center: $\mathbb{Z}_{2^n}$ wraps copies of $SU$ along the $U(1)$ component in a discontinuous way, and the quotient product $\times_{\mathbb{Z}_{2^n}}$ ``glues'' these copies into the global structure. Geometrically, the global structure is encoded in the sequence~\cite{nakahara2018geometry}:
$$
SU(2^n)\;\hookrightarrow\;U(2^n)\xrightarrow{\;\det\;} U(1)
$$
which defines $U(2^n)$ as a non-trivial principal $SU(2^n)$-bundle over the basis $U(1)$, induced by the determinant map. Each phase in $U(1)$ defines a local copy of $SU(2^n)$; as the phase varies around the circle, the fibers are glued together in a non-trivial way through the action of the discrete center $\mathbb{Z}_{2^n}$.

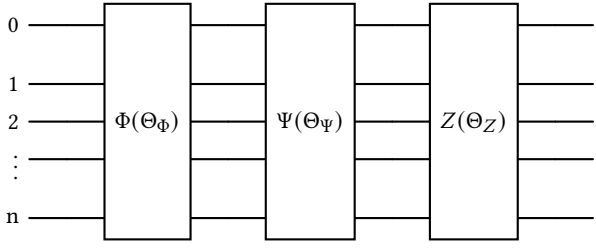
\begin{figure}[ht]
        \centering
        \caption{VQC for approximating $SU(2^n)$ operators with one single layer.}
            \begin{quantikz}
            \lstick{0}&&\gate[5]{\Phi(\Theta_\Phi)}&&\gate[5]{\Psi(\Theta_\Psi)}&&\gate[5]{Z(\Theta_Z)}&&\\
            \lstick{1}&&&&&&&&\\
            \lstick{2}&&&&&&&&\\
            \lstick{\vdots}&&&&&&&&\\
            \lstick{n}&&&&&&&&
            \end{quantikz}
        \label{fig:qcircuit_approx}
    \end{figure}

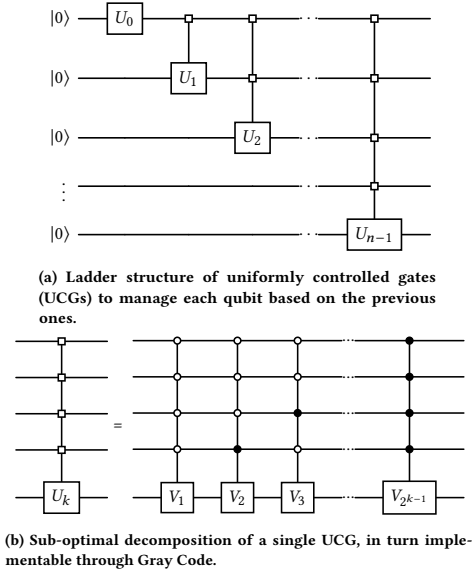
\begin{figure}[ht]
\centering
\caption{QSP natural framework.}
\resizebox{0.4\textwidth}{!}{
\subcaptionbox{Ladder structure of uniformly controlled gates (UCGs) to manage each qubit based on the previous ones.\label{qsp_structure}}{
\begin{quantikz}
    \lstick{$|0\rangle$}&\gate{U_0}&\octrl[style=rectangle]{1}&\octrl[style=rectangle]{1}&\push{\ldots}&\octrl[style=rectangle]{1}&\\
    \lstick{$|0\rangle$}&&\gate{U_1}&\octrl[style=rectangle]{1}&\push{\ldots}&\octrl[style=rectangle]{1}&\\
    \lstick{$|0\rangle$}&&&\gate{U_2}&\push{\ldots}&\octrl[style=rectangle]{1}&\\
    \lstick{$\vdots\mbox{ }$}&&&&\push{\ldots}&\octrl[style=rectangle]{1}&\\
    \lstick{$|0\rangle$}&&&&\push{\ldots}&\gate{U_{n-1}}&
\end{quantikz}}}\hfill
\resizebox{0.5\textwidth}{!}{
\subcaptionbox{Sub-optimal decomposition of a single UCG, in turn implementable through Gray Code.\label{multiplexor}}{
\begin{quantikz}
    &\octrl[style=rectangle]{1}&\\
    &\octrl[style=rectangle]{1}&\\
    &\octrl[style=rectangle]{1}&\\
    &\octrl[style=rectangle]{1}&\\
    &\gate{U_k}& 
\end{quantikz}=
\begin{quantikz}
    &\octrl{1}&\octrl{1}&\octrl{1}&\push{...}&\ctrl{1}&\\
    &\octrl{1}&\octrl{1}&\octrl{1}&\push{...}&\ctrl{1}&\\
    &\octrl{1}&\octrl{1}&\ctrl{1}&\push{...}&\ctrl{1}&\\
    &\octrl{1}&\ctrl{1}&\octrl{1}&\push{...}&\ctrl{1}&\\
    &\gate{V_1}&\gate{V_2}&\gate{V_3}&\push{...}&\gate{V_{2^{k-1}}}&
\end{quantikz}}}
\end{figure}

The SRBB elements $U_j^{(d)}$, where $1\leqslant j\leqslant 2^{2n}$, become the building blocks\footnote{The SRBB has $d^2=2^{2n}$ elements, but only the subset $\{U_j^{(d)}:\,1\leqslant j\leqslant d^2-1\}$ forms a basis for $\mathfrak{su}(d)$ when $d$ is even. This is because $U_{d^2}^{(d)}=\mathbb{I}_d$ for all $d$~\cite{belli2024scalable}.} of a unitary synthesis algorithm capable of approximating any given $U\in SU(2^n)$ according to~\cite{sarkar2023scalable}:
\begin{equation}\label{eqn:U_approx}
U_{approx}\equiv\prod_{l=1}^L\,Z(\Theta_Z^l)\Psi(\Theta_\Psi^l)\Phi(\Theta_\Phi^l)
\end{equation}
where $l$ is the layer index, indicating the number of repetitions of the approximating scheme depicted in Figure~\ref{fig:qcircuit_approx}. The SRBB component that represents the basis for $\mathfrak{su}(2^n)$ is partitioned into three subsets according to the algebraic properties of its matrix elements $U_j^{(d)}$ and is then used to define each macro-factor of equation~(\ref{eqn:U_approx}), which is in turn represented by the variational blocks of Figure~\ref{fig:qcircuit_approx}.
The QSP algorithm presented in this work only uses the diagonal $Z$-factor, defined by the following equation ($l=1$):
\begin{equation}
    Z(\Theta_Z)=\prod_{j=2}^{2^n}e^{i\left(\theta_{j^2-1}U_{j^2-1}^{(2^n)}\right)}
\end{equation}
where $U_{j^2-1}^{(2^n)}$ are diagonal SRBB elements and $\theta_{j^2-1}$ are the corresponding variational parameters. This factor can describe any \emph{diagonal} $SU(2^n)$ operator, thus any diagonal matrix of pure complex phases with a unit determinant. Defining the role of each macro-factor is useful for understanding why it is convenient to work with only the diagonal component of $\mathfrak{su}(2^n)$ in order to reduce the overall depth of the traditional QSP circuit (Figure~\ref{qsp_structure}) and approach the optimum achieved by~\cite{sun2023asymptotically}, but without the use of ancillae. The $\Psi$-factor is composed of SRBB elements which, by construction, already admit a $ZYZ$-decomposition (thus, they belong to the group $M_nZYZ$) and of other elements which admit the same decomposition only after the application of transposition matrices\footnote{Transpositions, or 2-cycles, are elements of the Permutation group $P_{2^n}$ with $2^n-2$ fixed points.} of even indices (which is why they are called even contributions). Therefore, the $\Psi$-factor can describe any $SU(2)$-block diagonal matrix belonging to $SU(2^n)$. Conversely, the $\Phi$-factor is made up of SRBB elements which admit a \emph{partial} $ZYZ$-decomposition only after the application of transposition matrices of odd indices (and so, they are called odd contributions). Therefore, the $\Phi$-factor can describe any $U(2)$-block diagonal matrix belonging to $SU(2^n)$. 

In \cite{10821064}, the SRBB synthesis algorithm was transferred into a quantum context through the scalable design of the corresponding VQC, and a new CNOT-reduced scalability scheme was identified and implemented with a single layer ($l=1$). Subsequently, the corresponding QNN was tested in both simulation and real quantum hardware~\cite{belli2024scalable}, to establish the effectiveness of the SRBB hierarchy in approximate synthesis tasks. The SRBB creates a direct connection between the topology of the special unitary group and the Lie algebra capable of representing unitary operators recursively, allowing the design of scalable variational circuits. How can this link be leveraged to design a state preparation algorithm with competitive depth trends while simultaneously defining a specific use case for SRBB? This question was the research motivation for this work.

According to the QSP paradigm, if the starting algebraic basis were complete SRBB, any unitary (and therefore any state) could be represented parametrically. However, thanks to the local group structure defined by Eq.~(\ref{eqn:local_structure}), it is sufficient to construct a parametric ansatz describing the arbitrary $SU(2^n)$ operator. In fact, the final state would deviate from the desired one only for a global phase, irrelevant to the experimental setup (real hardware) or easily implementable with a few single-qubit gates\footnote{A diagonal unitary matrix with entries equal to $e^{i\phi}$, for some $\phi\in\mathbb{R}$, can be added to the quantum circuit thanks to the combination $R_z(2\phi)\cdot XR(2\phi)X$, where $X$ is the Pauli gate and $R$ is the Phase Shift gate.}. At this point, if the natural framework for QSP is considered~\cite{mottonen2004transformation}, which involves a ladder structure of uniformly controlled gates (UCGs) throughout the quantum register as shown in Figure~\ref{qsp_structure}, each factor of the SRBB-based synthesis algorithm could have its own role.
In fact, in matrix form, UCG is a block-diagonal operator $U_k=diag(V_1,V_2,...,V_{2^{k-1}})\in\mathbb{C}^{2^k\times2^k}$, where each block $V_i\in U(2)$ is therefore representable through the $\Phi$-factor alone. In this case, for each UCG of the QSP circuit, a sub-circuit with $5\cdot2^{n-1}-6$ CNOT gates would be required~\cite{sarkar2024quantum}. If, instead, the UCG is restricted to the case of special unitary blocks $V_i\in SU(2^n)$, then it would be representable by the $\Psi$-factor alone, and the number of CNOTs for each controlled gate would go down to $3\cdot2^{n-1}-2$.

So long as the total number of CNOTs required in the $Z$-factor scaling scheme is $2^n-2$~\cite{belli2024scalable}, it would be much better to exploit only the diagonal sub-algebra and thus reduce the overall circuit depth. However, this approach is not compatible with the natural framework of QSP and the UCGs contained therein because the $Z$-factor parameterizes only diagonal unitary matrices, the entries of which are complex phases. One result can provide a variational approach to QSP that uses the SRBB diagonal component while maintaining the traditional structure: each unitary matrix $V_i\in U(2)$ can be decomposed in terms of $S$, $H$, and $R_z$ gates~\cite{sun2023asymptotically}, according to the well known identity
\begin{equation}\label{decomp_Ry}
R_y(\gamma_i)=SH\cdot R_z(\gamma_i)\cdot HS^\dag
\end{equation}
such that $V_i=e^{i\alpha_i}R_z(\beta_i)SHR_z(\gamma_i)HS^\dag R_z(\delta_i)$. In addition, the implementation of the $\Psi$- and $\Phi$-factors of SRBB exploits a Gray Code pattern equivalent to the decomposition of an arbitrary multiplexor or UCG (see Figure~\ref{multiplexor}), resulting in alternating sequences of CNOT, $R_z$, and $R_y$ gates~\cite{belli2024scalable}. Therefore, diagonal unitary matrices can encode the information of a UCG if their parameters are chosen appropriately, and with the help of single qubit gates like $S$ and Hadamard, a fact that allows the design of a scalable QSP circuit with only the diagonal component of the $\mathfrak{su}(2^n)$ Lie algebra.

\subsection{QSP natural structure}
\label{QSP}
\begin{figure}[ht]
\centering
\caption{Binary Search Tree for the complex amplitudes of the desired state.}
\begin{tikzpicture}[
    grow=down,
    level 1/.style={sibling distance=10em, level distance=4em},
    level 2/.style={sibling distance=6em, level distance=4em},
    edge from parent/.style={draw, ->, thick},
    edge label/.style={midway, sloped, above, font=\footnotesize, fill=white, text=black}
]
    \node {1}
        child { node {$\sqrt{0.3}$}
            child { node {$\sqrt{0.1}$}
                edge from parent node[edge label] {$|00\rangle$} }
            child { node {$\sqrt{0.2}$}
                edge from parent node[edge label] {$|01\rangle$} }
            edge from parent node[edge label] {$|0\cdots\rangle$} }
        child { node {$\sqrt{0.7}$}
            child { node {$\sqrt{0.4}$}
                edge from parent node[edge label] {$|10\rangle$} }
            child { node {$\sqrt{0.3}$}
                edge from parent node[edge label] {$|11\rangle$} }
            edge from parent node[edge label] {$|1\cdots\rangle$} };
\end{tikzpicture}
\label{bst}
\end{figure}
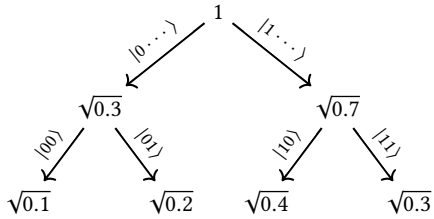

For a $n$-qubit system, a state preparation algorithm aims to prepare any desired quantum state $|\psi\rangle=\sum_{i=0}^{2^n-1}c_i|i\rangle$, where $c_i=|c_i|e^{i\phi_i}$ are the amplitudes associated with the vector
$c=(c_0,c_1,c_2,...,c_{2^n-1})\in\mathbb{C}^{2^n}$ with unit $l_2$-norm, eventually random. The QSP natural framework consists of a ladder structure of uniformly controlled gates (UCG), as illustrated in Figure~\ref{qsp_structure}, where each qubit $k$
is handled by a UCG that has the task of applying a single qubit unitary on it, conditioned on the basis state of
the first $k-1$ qubits. When defining the UCG to be included in the general structure, without losing generality, it is possible to consider $SU(2)$ blocks like $R_y(\theta)$, thus greatly simplifying the algorithm and its implementation. An exact QSP algorithm must find the best way to implement each UCG, since the usual decomposition of a multiplexor in the complete sequence of controlled gates, as shown in Figure~\ref{multiplexor}, allows for further implementation through a cyclic Gray Code and is suboptimal, as proven in~\cite{sun2023asymptotically}. Whether it is an exact or approximate QSP, the multiplexor structure in Figure~\ref{qsp_structure} requires a classical step to determine the UCG parameters necessary to achieve the desired state $|\psi\rangle$. The latter task can be performed thanks to the Binary Search Tree (BST), a diagram that stores the amplitudes $c_i$ of the desired state $|\psi\rangle$ in its leaf nodes (see Figure~\ref{bst}).
Every internal node stores the square root of the sum of squares of its child nodes, and the root node stores the
$l_2$-norm of the vector $c$. Moreover, it is necessary to consider the traditional decomposition of a UCG based on its own definition, illustrated in Figure~\ref{multiplexor}. In the following, the $n=2$ case exemplifies this classical step, which characterizes the natural structure of the QSP problem.

\subsubsection{Finding parameters for $n=2$}\label{sec:finding_params}
\begin{figure}[ht]
\centering
\caption{QSP traditional circuit for $n=2$.}
\begin{quantikz}
    \lstick{$|0\rangle$}&\gate{R_y(2\theta_0)}&\octrl{1}&\ctrl{1}&\\
    \lstick{$|0\rangle$}&&\gate{R_y(2\theta_1)}&\gate{R_y(2\theta_2)}&
\end{quantikz}
\label{qsp_n2}
\end{figure}
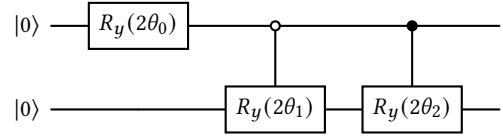

According to the natural framework, the fully decomposed quantum circuit to prepare an arbitrary state with 2 qubits is depicted in Figure~\ref{qsp_n2}. As anticipated, the choice of $R_y$-blocks allows us to simplify the classical determination of the natural parameters (the ones associated with the natural decomposition of Figure~\ref{multiplexor}). While in exact QSP this step is mandatory because these parameters must coincide with the new ones of the equivalent, and preferably better implemented, circuit (see, for instance,~\cite{sun2023asymptotically}), in its relaxed version, they represent the ideal target set of parameters that the variational approach must approximate. 
With reference to the BST illustrated in Figure~\ref{bst}, the parameter $\theta_0$ associated with $U_0=R_y(2\theta_0)$ is derived considering that $U_0$ acts only on the first qubit initialized in $|0\rangle$. Thus, $\theta_0=\arccos{\sqrt{\frac{0.3}{1}}}$. Similarly, the parameters $\theta_1$ and $\theta_2$ are derived considering that $U_1$ and $U_1$ act, respectively, when the quantum state of the register is $|00\rangle$ or $|10\rangle$, leading to $\theta_1=\arccos{\sqrt{\frac{0.1}{0.3}}}$ and $\theta_0=\arccos{\sqrt{\frac{0.4}{0.7}}}$. Obviously, not all nodes of the BST (and, therefore, not all amplitudes) influence the circuit parameters; some of them are indirectly considered by the normalization condition present at each level of the BST. In a matrix representation, the quantum circuit is equivalent to:
\begin{equation}\label{param_n2}
\small
\begin{split}    &\mbox{UCG}_2(2\theta_1,2\theta_2)\cdot[R_y(2\theta_0)\otimes\mathbb{I}_2]|0\rangle^{\otimes 2}=\\
    &=\begin{pmatrix}
        \cos{\theta_1}\cos{\theta_0}&-\sin{\theta_1}\cos{\theta_0}&-\cos{\theta_1}\sin{\theta_0}&\sin{\theta_1}\sin{\theta_0}\\
        \sin{\theta_1}\cos{\theta_0}&\cos{\theta_1}\cos{\theta_0}&-\sin{\theta_1}\sin{\theta_0}&-\cos{\theta_1}\sin{\theta_0}\\
        \cos{\theta_2}\sin{\theta_0}&-\sin{\theta_2}\sin{\theta_0}&\cos{\theta_2}\cos{\theta_0}&-\sin{\theta_2}\cos{\theta_0}\\
        \sin{\theta_2}\sin{\theta_0}&\cos{\theta_2}\sin{\theta_0}&\sin{\theta_2}\cos{\theta_0}&\cos{\theta_2}\cos{\theta_0}
    \end{pmatrix}
    \begin{pmatrix}
        1\\0\\0\\0
    \end{pmatrix}=\\
    &=\begin{pmatrix}
        \cos{\theta_1}\cos{\theta_0}\\\sin{\theta_1}\cos{\theta_0}\\\cos{\theta_2}\sin{\theta_0}\\\sin{\theta_2}\sin{\theta_0}
    \end{pmatrix}=
    \begin{pmatrix}
        \sqrt{0.1}\\\sqrt{0.2}\\\sqrt{0.4}\\\sqrt{0.3}
    \end{pmatrix}
\end{split}
\end{equation}
The choice $V_i=R_y(2\theta_i)$ simplifies the determination of the parameters of the quantum circuit; however, as shown by equation~\ref{param_n2}, it produces a completely real output vector whose components correspond to $|c_i|$, the modulus of the complex amplitudes of the desired vector $|\psi\rangle$. Indeed, from an experimental point of view (real hardware), a QSP algorithm must determine the probability distribution of the basis states of the quantum register corresponding to the desired quantum state, leaving out the information contained in the phases that is destroyed by the measurement process. However, by extending the domain of applicability of a QSP algorithm to the \emph{Encoding} process, it is important to also take into account the whole complex amplitude $c_i$, at least when the output of a QSP circuit is the input of another algorithm. In the next section, a part of the variational quantum circuit will be dedicated to that, and once again, only the $Z$-factor of the SRBB will be needed.

\subsection{VQC for the diagonal SRBB hierarchy}
\label{Z-factor}
\begin{figure*}[ht]
\centering
\caption{The $Z(\Theta_Z)$-factor circuit for $n=3$ with reduced CNOT-count.}
\includegraphics[width=0.7\textwidth]{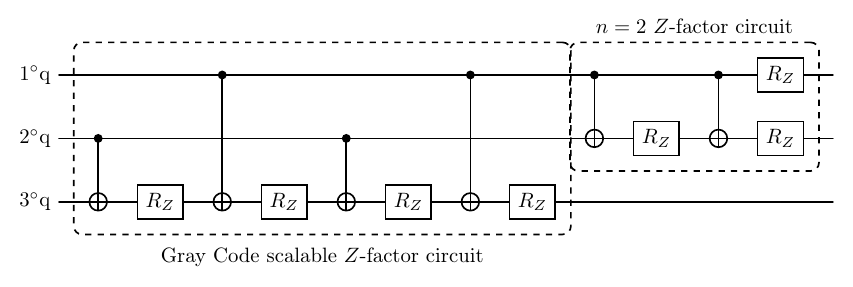}
\label{fig_Zsimpli_n3}
\end{figure*}

In~\cite{10821064}, a novel scalability scheme optimized in terms of CNOT gates is obtained for the diagonal components of the SRBB-based matrix algebra. As a result, the $Z$-factor can be easily implemented thanks to a partially recursive structure made up of two main blocks in which CNOTs and $R_z$ gates alternate. The circuit-level $Z$-factor construction algorithm involves sorting the diagonal elements of the algebra into a matrix of binary strings that represent the numbers from 1 to $2^n-1$. For each $n$, the aforementioned binary table presents some new algebraic contributions of the case under consideration (binary strings ending with 1) and others from the case n-1 (binary strings ending with 0). It has been demonstrated~\cite{belli2024scalable} that the new contributions of the case $n$ can be ordered to minimize the number of CNOTs via cyclic Gray Code matrices, while the old ones can be incorporated into the implementation of the case $n-1$. This is true except for the $n=2$ case, which falls outside the scalability scheme as the minimum number of qubits required to have non-trivial SRBB and, therefore, the starting point of the recursion. The sequence of rotation angles is defined by the same Gray Code matrix that encodes the optimization of the CNOTs. A detailed analysis of the CNOT optimization and the complete description of the resulting scalable algorithm for the $Z$-factor can be found in~\cite{belli2024scalable}; Figure~\ref{fig_Zsimpli_n3} exemplifies this scalable structure for the $n=3$ case.

\subsubsection{Implementing $R_y$-block UCGs}
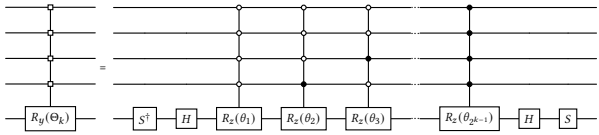
\begin{figure*}[ht]
\centering
\caption{$R_y$-block UCG circuit.}
\resizebox{0.65\textwidth}{!}{
\begin{quantikz}
    &\octrl[style=rectangle]{1}&\\
    &\octrl[style=rectangle]{1}&\\
    &\octrl[style=rectangle]{1}&\\
    &\octrl[style=rectangle]{1}&\\
    &\gate{R_y(\Theta_k)}& 
\end{quantikz}=
\begin{quantikz}
    &&&\octrl{1}&\octrl{1}&\octrl{1}&\push{...}&\ctrl{1}&&&\\
    &&&\octrl{1}&\octrl{1}&\octrl{1}&\push{...}&\ctrl{1}&&&\\
    &&&\octrl{1}&\octrl{1}&\ctrl{1}&\push{...}&\ctrl{1}&&&\\
    &&&\octrl{1}&\ctrl{1}&\octrl{1}&\push{...}&\ctrl{1}&&&\\
    &\gate{S^\dag}&\gate{H}&\gate{R_z(\theta_1)}&\gate{R_z(\theta_2)}&\gate{R_z(\theta_3)}&\push{...}&\gate{R_z(\theta_{2^{k-1}})}&\gate{H}&\gate{S}&
\end{quantikz}}
\label{decomp_Ry_circuit}
\end{figure*}

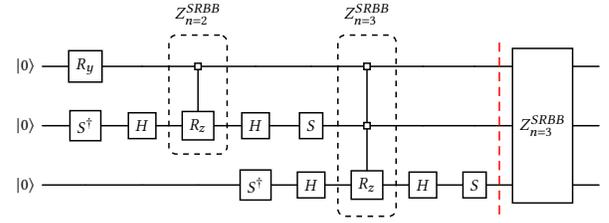
\begin{figure*}[ht]
\centering
\caption{QSP circuit with insertions for the $Z$-factor in the $n=3$ case. The barrier separates the component that learns the moduli from the one that learns the phases.}
\resizebox{0.65\textwidth}{!}{
\begin{quantikz}
    \lstick{$|0\rangle$}&\gate{R_y}&&\octrl[style=rectangle]{1}\gategroup[2,steps=1,style={dashed,rounded corners},label style={label position=above,anchor=north,yshift=+0.5cm}]{$Z^{SRBB}_{n=2}$}&&&\octrl[style=rectangle]{1}\gategroup[3,steps=1,style={dashed,rounded corners},label style={label position=above,anchor=north,yshift=+0.5cm}]{$Z^{SRBB}_{n=3}$}&&\slice{}&\gate[3]{Z^{SRBB}_{n=3}}&\\
    \lstick{$|0\rangle$}&\gate{S^\dag}&\gate{H}&\gate{R_z}&\gate{H}&\gate{S}&\octrl[style=rectangle]{1}&&&&\\
    \lstick{$|0\rangle$}&&&&\gate{S^\dag}&\gate{H}&\gate{R_z}&\gate{H}&\gate{S}&&
\end{quantikz}
}
\label{qsp_n3}
\end{figure*}

According to identity~(\ref{decomp_Ry}), an arbitrary UCG $U_k$ composed only of $R_y$ blocks and with $\Theta_k=(\theta_1,\theta_2,...,\theta_{2^{k-1}})$ admits the following decomposition in terms of multi-controlled $R_z$-gates:
\begin{equation}\label{eqn:multiplex_reduced}
\begin{split}
U_k&=[\mathbb{I}_{n-1}\otimes(SH)]\cdot\\
&\cdot\mbox{diag}[R_z(\theta_1),R_z(\theta_2),...,R_z(\theta_{2^{k-1}})]\cdot\\
&\cdot[\mathbb{I}_{n-1}\otimes(HS^\dag)]
\end{split}
\end{equation}
in such a way that, at the circuit level, the substitution illustrated in Figure~\ref{decomp_Ry_circuit} occurs.
Therefore, the QSP ladder structure of Figure~\ref{qsp_structure} can be redesigned with only gates $S$, $H$ and $R_z$, a configuration that is formally equivalent to a sequence of diagonal unitary matrices of pure complex phases and thus representable through the diagonal component of SRBB. Figure~\ref{qsp_n3} shows the case $n=3$, where the circuit blocks that can be approximated by the $Z$-factor of the SRBB are highlighted with a dotted line. A final circuit block consisting of $Z^{SRBB}_{n_{max}}$ has been added to the original scale structure and is separated by a dotted line; the latter encodes the necessary phase correction after choosing to consider only the $R_y$ blocks in the UCGs, as explained in Section~\ref{sec:finding_params}.

\subsubsection{Depth and gate counts}
The depth trend and gate count as a function of the number of qubits $n$ can easily be deduced by following the recursive structure given by the $Z$-factor in the overall circuit.

Since the recursive structure is valid from $n\geqslant3$, we set aside the contribution to the depth given by the $R_y$ on the first qubit and by the first UCG$_2$\footnote{The symbol UCG$_n$ indicates the multiplexor upon $n$ qubits.}, a contribution which is equal to $2+Z^{SRBB}_{n=2}+2=8$. As can be seen in Figure~\ref{qsp_n3}, at every $n\geqslant3$ the $S^\dag$ and $H$ gates do not create depth as they are parallel to the circuit for $Z^{SRBB}_{n-1}$. The same occurs for the gates $H$ and $S$ following the implementation of $Z^{SRBB}_n$. This implies that from $n\geqslant3$, the depth is determined only by the $Z$-factors up to $n_{max}$, as stated by the following equation:
\begin{equation}
    D(n_{max})=8+\sum_{n=3}^{n_{max}}\mathcal{Z}_n+\mathcal{Z}_{n_{max}}
\end{equation}
where $\mathcal{Z}_n$ represents the contribution of each $Z^{SRBB}_n$ factor according to
\begin{equation}
    \mathcal{Z}_n=\sum_{i=3}^{n}(2^i-1)+4=2^{n+1}-2-n
\end{equation}
Thus, for $n_{max}=M$, it holds that
\begin{equation}
    \begin{split}
        D(M)&=8+\sum_{n=3}^{M}(2^{n+1}-2-n)+2^{M+1}-2-M=\\
        &=2^{M+2}-2M-4-\Delta_{3,M}+2^{M+1}-2-M
    \end{split}
\end{equation}
where $\Delta_{3,M}$ represents the sum of the integers from 3 to $M$, obtained from the well known Euler's formula $S_{n_1\rightarrow n_2}=\frac{n_2(n_2+1)-n_1(n_1+1)}{2}+n_1$, for $n_1<n_2$. Substituting into the previous equation, the depth trend as a function of the number of qubits $n$ becomes:
\begin{equation}
    D(n)=6\cdot2^n-\frac{n^2+7n}{2}-3
\end{equation}
Finally, for each $Z^{SRBB}_n$ for $n\geqslant3$, the number of CNOTs is equal to $2^n-2$, while the number of $R_z$ is equal to $2^n-1$. Following a similar reasoning to that of depth, the total number of rotations is:
\begin{equation}
\begin{split}
    N_{rot}(n)&=4+\mathcal{Z}_n^{rot}+2^n-1=\\
    &=3\cdot2^n-n-3
\end{split}
\end{equation}
where
\begin{equation}
   \mathcal{Z}_n^{rot}=\sum_{i=3}^{n}(2^i-1)=2^{n+1}-n-6 
\end{equation}
The total number of CNOT gates can be computed as follows:
\begin{equation}
\begin{split}
    N_{CNOT}(n)&=2+\mathcal{Z}_n^{CNOT}+2^n-2=\\
    &=3\cdot2^n-2n-4
\end{split}
\end{equation}
where
\begin{equation}
   \mathcal{Z}_n^{CNOT}=\sum_{i=3}^{n}(2^i-2)=2^{n+1}-2n-4 
\end{equation}

\section{QNN Implementation and Results}
\label{lab:results}
The proposed VQC has been implemented using the PennyLane library~\cite{bergholm2018pennylane}. The tests have been performed using the simulator provided by the library, executed on a Linux machine equipped with an AMD EPYC 7282 CPU and 256 GB of RAM. Moreover, the VQC is tested on real quantum computers provided by IQM, namely IQM Deneb and IQM Garnet, as well as on devices accessed through the AWS service, specifically Rigetti Ankaa-3 and IonQ Forte1.

\begin{figure*}[ht]
\centering
\caption{General structure of the QNN, divided in the two component: $SU_{Modulus}$, which learns the moduli of the state, and $SU_{Phase}$, which learns the phases.}
\resizebox{0.55\textwidth}{!}{
\begin{quantikz}
    \lstick{$|0\rangle$}&\gate[4]{VQC_{QSP}^{SRBB} = SU_{Modulus}}&\gate[4]{Z^{SRBB}_{n} = SU_{Phase}}&\\
    \lstick{$|0\rangle$}&&&\\
    \lstick{$\vdots\mbox{ }$}&&&\\
    \lstick{$|0\rangle$}&&&
\end{quantikz}
}
\label{qnn}
\end{figure*}
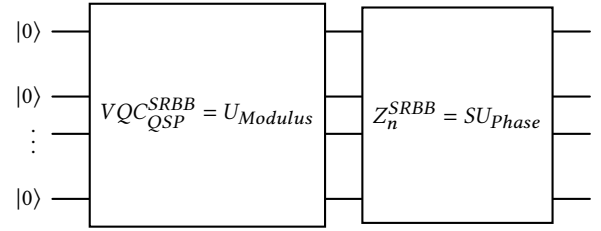

Figure~\ref{qnn} provides a general depiction of the QNN, which is used to prepare the desired state and operates in two steps, as indicated by the barrier in Figure~\ref{qsp_n3} with 3 qubits:
\begin{enumerate}
    \item the first part of the network learns the modulus of each component of the desired state;
    \item the second part of the network learns how to produce the phase for each component of the desired state.
\end{enumerate}
When the VQC has to learn the moduli of the state, the special unitary $SU_{Modulus}$ associated with the QSP natural structure is calculated following Section \ref{QSP} and then approximated by the first part of the circuit in Figure~\ref{qnn}. On the other hand, when the VQC has to learn the phases, the target unitary $U_{phase}$ is a diagonal matrix, whose diagonal elements are defined as the complex phases of each amplitude of the desired state. Subsequently, the $2^n$ $SU_{phase}$ matrices related to $U_{phase}$ must be calculated, as the Z-factor employed in the implementation of the second part of the circuit in Figure~\ref{qnn} can only represent $SU$ operators~(Section \ref{Z-factor}), so it must learn one of these $SU$, using the same technique introduced in~\cite[Section 5]{belli2024scalable}.
The training of these two components cannot be simultaneous, as it would prevent the network from distinguishing which part is responsible for each task, making it a more difficult challenge and, therefore, hindering the network's ability to correctly learn the circuit.

At the end of the training, the resulting network generates the exact state up to a global phase, which arises from the approximation of both the modulus and the phase block circuits.
In any case, this issue can be corrected in $U_{phase}$ by multiplying the unitary matrix by the root of the determinant of the corresponding $SU$, as explained in Section~\ref{lab:theoretical_framewowk}.
However, different considerations are required when addressing the modulus circuit. During the approximation of $SU_{Modulus}$ with fidelity or trace distance as losses, some interesting features of the chosen decomposition and \emph{group structure} emerge in the complex component (including global phase) of the approximate operator. Although the circuit block $VQC_{QSP}^{SRBB}$ in Figure~\ref{qnn} is designed to prepare only the real part of the desired state via a sequence of $UCR_y$ that cannot leave the $SU(2^n)$ subgroup\footnote{The ladder structure of $UCR_y$, rewritten by means of Eq.~(\ref{eqn:multiplex_reduced}), is implemented via the diagonal component of SRBB, which collects generators all belonging to the $\mathfrak{su}(2^n)$ Lie algebra.}, in some cases, the approximated matrix shows a non-zero imaginary part (unless numerical approximations are used) and a global phase.
The imaginary part can be explained simply by thinking of the parametric matrix corresponding to the $Z$-factor, that is, a diagonal matrix of complex phases with arguments being linear combinations of the parameters~\cite{10821064}, and the QNN's attempt to approximate an ideal multiplexor (see, for example, Figure~\ref{qsp_n3}). Clearly, in the approximation process, the S- and H-gates fail to perfectly combine the phases and reproduce $UCR_y$. The global phase emerges instead for reasons attributable to the \emph{global structure} of the $U(2^n)$ group\footnote{See Section~\ref{lab:theoretical_framewowk} for a formal discussion of the non-trivial phase bundle structure due to the discrete center of $SU(N)$ isomorphic to $\mathbb{Z}_N$.}. On the one hand, the optimization of observables such as trace and fidelity, which are global phase invariants, induces a \emph{residual} $U(1)$ symmetry on the parametric landscape; on the other hand, the membership group of $VQC_{QSP}^{SRBB}$ in Figure~\ref{qnn}, which is $SU(2^n)$, admits a subset of \emph{discrete} global phases (discrete center of $SU(2^n)$ isomorphic to $\mathbb{Z}_{2^n}$) that preserves the unit determinant and induces a \emph{topological quantization} of the phase. This can be clearly seen from:
$$
\det(e^{i\phi}S)=e^{i\phi d}\det(V)=e^{i\phi d}=1\quad\leftrightarrow\quad\phi=\frac{2\pi k}{d}
$$
where $S\in SU(2^n)$, $d=2^n$, and $k\in\mathbb{Z}$. Therefore, the ansatz for $VQC_{QSP}^{SRBB}$ cannot generate an arbitrary complex phase, but only the discrete complex phases that preserve $SU$ and these are precisely the integer multiples of $\frac{2\pi}{d}$. Therefore, for example, in the case of the Bell state $\ket{\Phi^{+}} = \frac{\ket{00} + \ket{11}}{\sqrt{2}}$, there are four possible global phases corresponding to the multiples of $\frac{2\pi}{d} = \frac{\pi}{2}$, namely $0, \frac{\pi}{2}, \pi, $ and $\frac{3\pi}{2}$. These generate the following four equivalent states: 
\begin{enumerate}
    \item $[\frac{1}{\sqrt{2}} + 0.0j, 0.0 + 0.0j, 0.0 + 0.0j, \frac{1}{\sqrt{2}} + 0.0j]$
    \item $[0.0 + \frac{1}{\sqrt{2}}j, 0.0 + 0.0j, 0.0 + 0.0j, 0.0 + \frac{1}{\sqrt{2}}j]$
    \item $[-\frac{1}{\sqrt{2}} + 0.0j, 0.0 + 0.0j, 0.0 + 0.0j, -\frac{1}{\sqrt{2}} + 0.0j]$
    \item $[0.0 - \frac{1}{\sqrt{2}}j, 0.0 + 0.0j, 0.0 + 0.0j, 0.0 + -\frac{1}{\sqrt{2}}j]$
\end{enumerate}
Finally, from the optimizer's point of view, the \emph{continuous} degeneracy of minima induced by the $U(1)$ symmetry coexists with the $SU$-preserving discrete phases, since a phase-invariant loss function formally extends the discrete center $\mathbb{Z}_{2^n}$ to the effective $U(1)$. In other words, the gradient is zero along the direction $U\rightarrow e^{i\phi}U$, even though the ansatz cannot follow that infinitesimal direction except by hopping between discrete points of the center. In the language of Section~\ref{lab:theoretical_framewowk}, this same fact is formulated by stating that the optimizer can converge to any $SU$ in the phase bundle above the target. The set of $SU$-preserving phases is discrete, not because of choices made by the optimizer, but because it is imposed by the underlying topological connection.
Alternatively, if needed, the total global phase, given by the sum of the phases produced by the two components of the VQC, can be corrected directly within the quantum circuit. The global phase can be added to the diagonal elements by applying a unitary operation to one qubit, consisting of the gate sequence $R_z(2\phi)\cdot PauliX\cdot \text{PhaseShift}(2\phi)\cdot PauliX$, where $\phi$ is the global phase and:
\begin{equation}
    \text{PhaseShift}(\phi) = \begin{pmatrix} 
1 & 0 \\
0 & e^{i\phi} 
\end{pmatrix}
\end{equation}
\begin{equation}
R_z(\phi) = \begin{pmatrix} 
e^{-i\frac{\phi}{2}} & 0 \\
0 & e^{i\frac{\phi}{2}} 
\end{pmatrix}
\end{equation}

All results from the real quantum computers (Section~\ref{IQM}) have been obtained using the original circuit, without correcting the total global phase, as the objective is to assess the algorithm's performance. Moreover, the error has been computed by calculating the Hellinger distance between the target probabilities and those produced by the network. Therefore, the global phase does not affect the results. 

Different loss functions are tested:
\begin{itemize}
    \item Frobenius norm: $\|A \|_{F} = \sqrt{\operatorname{tr}(A A^\dagger)}$, where $A = SU_{ideal} - SU_{VQC}$
    \item Trace distance: $
            \| \rho - \sigma \|_1 = \operatorname{tr}\sqrt{(\rho - \sigma)^\dagger(\rho - \sigma)}$
    \item Fidelity: $
            \operatorname{F}(\rho, \sigma) = \left(\operatorname{tr}\sqrt{\sqrt{\rho}\sigma\sqrt{\rho}}\right)^2$, where $\rho$ and $\sigma$ are the density matrices of the states obtained by the ideal $SU$ or $U$ operation, which will be approximated, and by the VQC, respectively, given the same input state.
\end{itemize}

When the Frobenius norm is used as the loss function, the VQC is designed to obtain its matrix representation. The Frobenius distance is then calculated to evaluate the loss. This approach involves comparing the $SU(2^n)_{Phase}$ or $SU(2^n)_{Modulus}$ matrix associated with the target circuit that the QNN aims to approximate and the $SU(2^n)_{Phase}$ or $SU(2^n)_{Modulus}$ matrix associated with the VQC.

On the other hand, when either fidelity or trace distance is used as the loss function, a training set of 1000 random states is generated. The loss function is minimized by adjusting the VQC to decrease the trace distance (or increase the fidelity) between the states produced by the ideal matrix and those obtained from the approximated matrix.

Two different optimizers are used based on the chosen loss function:
\begin{itemize}
    \item Adam~\cite{kingma2014adam}: used with trace and fidelity, with a learning rate $0.01$, $50$ epochs, and a batch size $64$; 
    \item Nelder Mead~\cite{nelder1965simplex}: used with Frobenius, with $10^{-15}$ as the objective error for 2 and 3 qubits, while $10^{-10}$ applies to 4 qubits; otherwise, the training time becomes excessively long.
\end{itemize}

\subsection{Quantum Circuit Simulation}
The QNN has been tested with different numbers of qubits, ranging from 2 to 8, using both random states and specific states, such as Bell, GHZ, and uniform superposition. The mean results are shown in Table~\ref{tab:simulationResults}. The error is computed as the trace distance between the target state and the state achieved by the network.

The Nelder Mead optimizer demonstrates effective performance with up to 4 qubits. However, when the circuit has to approximate 4-qubit states, the accepted error has been decreased from $10^{-15}$ to $10^{-10}$ to achieve the approximation in an acceptable time, as shown in the table. As the number of qubits increases, the problem becomes more complex, and the optimizer cannot converge effectively.

On the other hand, the Adam optimizer exhibits longer training times for fewer qubits, but these do not increase exponentially. Furthermore, when fidelity is used as a loss function, the network's results closely match those achieved with Nelder Mead. Moreover, the network can also approximate states with higher amplitudes. Conversely, using trace distance as the loss function significantly diminishes the network's performance compared to the other metrics.

\begin{table}[htbp]
\centering
\renewcommand{\arraystretch}{1.2}
\caption{Results achieved in simulation from 2 to 8 qubits.}
    \begin{tabular}{|>{\centering\arraybackslash}m{0.6cm}||>{\centering\arraybackslash}m{1cm}|>{\centering\arraybackslash}m{1.2cm}|>{\centering\arraybackslash}m{1.2cm}|>{\centering\arraybackslash}m{1cm}|>{\centering\arraybackslash}m{1.2cm}|}
     \hline
     n qubit & Time taken with Adam & Error with Adam + fidelity & Error with Adam + trace distance & Time taken with Nelder Mead & Error with Nelder Mead  \\
     \hline
     2 & $\sim$~$9m$ & $\sim$~$10^{-15}$ & $\sim$~$10^{-3}$ & $\sim$~$2s$ & $\sim$~$10^{-15}$\\
     \hline
     3 & $\sim$~$16m$ & $\sim$~$10^{-13}$ & $\sim$~$10^{-3}$ & $\sim$~$30s$ & $\sim$~$10^{-15}$\\
     \hline
     4 & $\sim$~$33m$ & $\sim$~$10^{-13}$ & $\sim$~$10^{-3}$ & $\sim$~$9m$ & $\sim$~$10^{-10}$\\
     \hline
     5 & $\sim$~$1h$ & $\sim$~$10^{-7}$ & $\sim$~$10^{-3}$ & $-$ & $-$\\
     \hline
     6 & $\sim$~$2h$ & $\sim$~$10^{-3}$ & $\sim$~$10^{-3}$ & $-$ & $-$\\
     \hline
     7 & $\sim$~$5h$ & $\sim$~$10^{-3}$ & $\sim$~$10^{-3}$ & $-$ & $-$\\
     \hline
     8 & $\sim$~$11h$ & $\sim$~$10^{-3}$ & $\sim$~$10^{-3}$ & $-$ & $-$\\
     \hline
    \end{tabular}
    
    \label{tab:simulationResults}
\end{table}

As shown in Table~\ref{tab:0results}, the network can reproduce the identity, as it can reproduce \ket{0} with high precision. Since the results are equal to the random states, only the results from 2 to 5 qubits are presented. In addition, in Fig.~\ref{fig:identity}, the approximated unitary is presented for the 2-qubit circuit.

\begin{table}[htbp]
\centering
\renewcommand{\arraystretch}{1.2}
\caption{Results achieved in simulation when reproducing the \ket{0} state.}
    \begin{tabular}{|>{\centering\arraybackslash}m{1cm}||>{\centering\arraybackslash}m{2.5cm}|}
     \hline
     n qubit & Error with Adam + fidelity  \\
     \hline
     2 & $\sim$~$10^{-15}$ \\
     \hline
     3 & $\sim$~$10^{-13}$ \\
     \hline
     4 & $\sim$~$10^{-13}$ \\
     \hline
     5 & $\sim$~$10^{-7}$ \\
     \hline
    \end{tabular}
    
    \label{tab:0results}
\end{table}

\begin{figure*}[h!]
\centering
\caption{Produced matrix when approximating the $\ket{0}$ state.}
\scalebox{1}{
$
I + 10^{-15}
\begin{pmatrix}
-0.1 + 3.42865i & 2.27767 & 0.57779 & 0 \\
-2.27767 & -0.1 + 3.42865i & 0 & 0.57779 \\
-0.57779 & 0 & -0.1 - 3.42865i & -0.9917 \\
0 & -0.57779 & 0.9917 & -0.1 - 3.42865i
\end{pmatrix}
$
}
\label{fig:identity}
\end{figure*}

Moreover, the network is tested on preparing a random state from another random state, rather than only from \ket{0}, as shown in Table~\ref{tab:randomResultsSim}. To perform this state preparation, the circuit described in~\ref{qnn} is applied twice: first, to approximate the \ket{0} state, and then, starting from \ket{0}, to prepare the target random state. When the circuit is applied to prepare the \ket{0} state, the components (for example, the 3-qubit components in Fig.~\ref{fig_Zsimpli_n3}) of the two unitaries in~\ref{qnn} are applied in inverse order, following the approach in~\cite{mottonen2004transformation}.

\begin{table}[htbp]
\centering
\renewcommand{\arraystretch}{1.2}
\caption{Results achieved in simulation from 2 to 8 qubits, starting from random states.}
    \begin{tabular}{|>{\centering\arraybackslash}m{1cm}||>{\centering\arraybackslash}m{2.5cm}|>{\centering\arraybackslash}m{2.5cm}|}
     \hline
     n qubit & Error with Adam + fidelity & Error with Nelder Mead  \\
     \hline
     2 & $\sim$~$10^{-14}$ & $\sim$~$10^{-15}$\\
     \hline
     3 & $\sim$~$10^{-13}$ & $\sim$~$10^{-15}$\\
     \hline
     4 & $\sim$~$10^{-13}$ & $-$\\
     \hline
     5 & $\sim$~$10^{-4}$ & $-$\\
     \hline
     6 & $\sim$~$10^{-3}$ & $-$\\
     \hline
     7 & $\sim$~$10^{-3}$ & $-$\\
     \hline
    \end{tabular}
    
    \label{tab:randomResultsSim}
\end{table}

\subsection{Real Quantum Computers}
\label{IQM}
This section presents the results obtained from running the trained VQC (with fidelity as the loss function) on real quantum computers. Specifically, the experiment was carried out using quantum computers provided by the AWS service and two additional quantum computers provided by IQM:
\begin{itemize}
    \item IQM Garnet. It is a 20-qubit quantum processing unit based on superconducting transmon qubits. The qubits are arranged in a square lattice and connected by tunable couplers. The system is calibrated to support arbitrary X and Y rotations as the native single-qubit gate and CZ as the native two-qubit gate~\cite{abdurakhimov2024technologyperformancebenchmarksiqms}.
    \item IQM Deneb. It is a 6-qubit quantum processing unit where all superconducting transmon qubits are connected to one central computational resonator. In addition to single qubit rotations, all qubits can perform CZ gates within the subspace formed by the two lowest energy levels of the resonator. Furthermore, a MOVE operation is calibrated to exchange the occupations of the qubit-resonator states 0-1 and 1-0. It can, therefore, be used to shuttle excitations back and forth between any qubit and the resonator~\cite{PhysRevResearch.4.043089}.
\end{itemize}

For each number of qubits, $50$ random states have been tested, and the results are presented in Table~\ref{tab:IQMResults} and Table~\ref{tab:IQMGarnetResults}. In this case, the Hellinger distance between the target probabilities and those achieved by the VQC is computed as the error. With 2 and 3 qubits, the results achieved are quite good; however, as the number of qubits increases, the performance declines due to the growing complexity of the VQC. However, with IQM Garnet, the error does not increase significantly, and the performance remains good even with more qubits.

\begin{table}[htbp]
\centering
\renewcommand{\arraystretch}{1.2}
\caption{Results achieved on IQM Deneb from 2 to 5 qubits.}
    \begin{tabular}{|>{\centering\arraybackslash}m{1cm}||>{\centering\arraybackslash}m{2cm}|>{\centering\arraybackslash}m{2cm}|}
     \hline
     n qubit & mean Hellinger distance & std Hellinger distance  \\
     \hline
     2 & $\sim$~$0.06$ & $\sim$~$0.03$ \\
     \hline
     3 & $\sim$~$0.19$ & $\sim$~$0.09$ \\
     \hline
     4 & $\sim$~$0.36$ & $\sim$~$0.05$ \\
     \hline
     5 & $\sim$~$0.51$ & $\sim$~$0.06$ \\
     \hline
    \end{tabular}
    
    \label{tab:IQMResults}
\end{table}

\begin{table}[htbp]
\centering
\renewcommand{\arraystretch}{1.2}
\caption{Results achieved on IQM Garnet from 2 to 5 qubits.} 
    \begin{tabular}{|>{\centering\arraybackslash}m{1cm}||>{\centering\arraybackslash}m{2cm}|>{\centering\arraybackslash}m{2cm}|}
     \hline
     n qubit & mean Hellinger distance & std Hellinger distance  \\
     \hline
     2 & $\sim$~$0.03$ & $\sim$~$0.02$ \\
     \hline
     3 & $\sim$~$0.11$ & $\sim$~$0.08$ \\
     \hline
     4 & $\sim$~$0.19$ & $\sim$~$0.04$ \\
     \hline
     5 & $\sim$~$0.31$ & $\sim$~$0.05$ \\
     \hline
    \end{tabular}
    
    \label{tab:IQMGarnetResults}
\end{table}

In addition, the network has been tested with specific states. The results are presented in Table~\ref{tab:IQMResults2Qubits} for 2-qubit states, Table~\ref{tab:IQMResults3Qubits} for 3-qubit states, and Table~\ref{tab:IQMResults4Qubits} for 4-qubit states. The results demonstrate that the network can accurately approximate states such as uniform superposition, Bell, and GHZ. However, in some cases, the quality of the approximation significantly decreases. 

\begin{table}[htbp]
\centering
\renewcommand{\arraystretch}{1.2}
\caption{Results obtained on IQM Deneb and IQM Garnet using 2 qubits with sparse state or state with a uniform superposition.}
    \begin{tabular}{|>{\centering\arraybackslash}m{3cm}||>{\centering\arraybackslash}m{1.8cm}|>{\centering\arraybackslash}m{1.8cm}|}
     \hline
     state & IQM Deneb Hellinger distance & IQM Garnet Hellinger distance  \\
     \hline
     [$1/\sqrt{2}$, $0$, $0$, $1/\sqrt{2}$] or Bell & $\sim$~$0.13$ & $\sim$~$0.12$  \\
     \hline
     [$1/\sqrt{2}$, $0$, $1/\sqrt{2}$, $0$]& $\sim$~$0.04$ & $\sim$~$0.09$ \\
     \hline
     [$0.5$, $0.5$, $0.5$, $0.5$] & $\sim$~$0.01$ & $\sim$~$0.01$ \\
     \hline
     [$0$, $0$, $1/\sqrt{2}$, $-1/\sqrt{2}$] & $\sim$~$0.15$ & $\sim$~$0.12$ \\
     \hline
     [$-1/\sqrt{2}$, $1/\sqrt{2}$, $0$, $0$] & $\sim$~$0.03$ & $\sim$~$0.1$ \\
     \hline
    \end{tabular}
    
    \label{tab:IQMResults2Qubits}
\end{table}

\begin{table}[htbp]
\centering
\renewcommand{\arraystretch}{1.2}
 \caption{Results obtained on IQM Deneb and IQM Garnet using 3 qubits with sparse state or state with a uniform superposition.}
    \begin{tabular}{|>{\centering\arraybackslash}m{4cm}||>{\centering\arraybackslash}m{1.8cm}|>{\centering\arraybackslash}m{1.8cm}|}
     \hline
     state & IQM Deneb Hellinger distance & IQM Garnet Hellinger distance  \\
     \hline
     [$1/\sqrt{2}$, $0$, $0$, $0$, $0$, $0$, $0$, $1/\sqrt{2}$] & $\sim$~$0.38$  & $\sim$~$0.31$ \\
     \hline
     [$0.35$, $0.35$, $0.35$, $0.35$, $0.35$, $0.35$, $0.35$, $0.35$] & $\sim$~$0.1$ & $\sim$~$0.06$ \\
     \hline
     [$1/\sqrt{2}$, $1/\sqrt{2}$, $0$, $0$, $0$, $0$, $0$, $0$] & $\sim$~$0.23$ & $\sim$~$0.23$ \\
     \hline
     [$0$, $0$, $0$, $0$, $0$, $0$, $1/\sqrt{2}$, -$1/\sqrt{2}$] & $\sim$~$0.46$ & $\sim$~$0.25$ \\
     \hline
     [$0$, $1/\sqrt{2}$, $0$, $0$, $0$, -$1/\sqrt{2}$, $0$, $0$] & $\sim$~$0.29$ & $\sim$~$0.26$ \\
     \hline
    \end{tabular}
   
    \label{tab:IQMResults3Qubits}
\end{table}

\begin{table}[htbp]
\centering
\renewcommand{\arraystretch}{1.2}
    \caption{Results obtained on IQM Deneb and IQM Garnet using 4 qubits with sparse state or state with a uniform superposition.}
    \begin{tabular}{|>{\centering\arraybackslash}m{4cm}||>{\centering\arraybackslash}m{1.8cm}|>{\centering\arraybackslash}m{1.8cm}|}
     \hline
     state & IQM Deneb Hellinger distance & IQM Garnet Hellinger distance \\
     \hline
     [$0.25$, $0.25$, $0.25$, $0.25$, $0.25$, $0.25$, $0.25$, $0.25$, $0.25$, $0.25$, $0.25$, $0.25$, $0.25$, $0.25$, $0.25$, $0.25$] & $\sim$~$0.27$ & $\sim$~$0.14$  \\
     \hline
     [$0$, $0$, $0$, $0$, $0$, $0$, $0$, $0$, $0$, $0$, $0$, $0$, $0$, $0$, $1/\sqrt{2}$, $-1/\sqrt{2}$] & $\sim$~$0.74$ & $\sim$~$0.56$ \\
     \hline
     [$-1/\sqrt{2}$, $1/\sqrt{2}$, $0$, $0$, $0$, $0$, $0$, $0$, $0$, $0$, $0$, $0$, $0$, $0$, $0$, $0$] & $\sim$~$0.33$ & $\sim$~$0.32$ \\
     \hline
    \end{tabular}
    
    \label{tab:IQMResults4Qubits}
\end{table}

Furthermore, the error produced by the network when approximating a target state from a random initial state is reported in Table~\ref{tab:RealHW}, with 1024 shots. The tests are carried out by applying the two pre-trained circuits to the \ket{0} state and comparing the obtained results with those from the simulation. The error is computed by using the Hellinger distance between the two probability distributions. The circuits are tested on Rigetti Ankaa-3, IQM Garnet, and IonQ Forte1 through AWS braket. The circuits are also tested with 4096 and 8192 shots on Rigetti Ankaa-3 in Table~\ref{tab:RealHW2} to see how the number of shots affects performance.

\begin{table}[htbp]
\centering
\renewcommand{\arraystretch}{1.2}
    \caption{Results achieved on real hardware when approximating a target state from a random initial state, with 1024 shots.} 
    \begin{tabular}{|>{\centering\arraybackslash}m{1cm}||>{\centering\arraybackslash}m{2cm}|>{\centering\arraybackslash}m{2cm}|>{\centering\arraybackslash}m{2cm}|}
     \hline
     n qubit & Rigetti Ankaa-3 & IQM Garnet & IonQ Forte1  \\
     \hline
     2 & $\sim$~$0.2$ & $\sim$~$0.06$ & $\sim$~$0.05$\\
     \hline
     3 & $\sim$~$0.6$ & $\sim$~$0.4$ & $\sim$~$0.09$\\
     \hline
     4 & $\sim$~$0.6$ & $\sim$~$0.44$ & $\sim$~$0.19$\\
     \hline
     5 & $\sim$~$0.77$ & $-$ & $\sim$~$0.5$\\
     \hline
    \end{tabular}
    \label{tab:RealHW}
\end{table}

\begin{table}[htbp]
\centering
\renewcommand{\arraystretch}{1.2}
    \caption{Results achieved on real hardware when approximating a target state from a random initial state, with 4096 and 8192 shots.} 
    \begin{tabular}{|>{\centering\arraybackslash}m{1cm}||>{\centering\arraybackslash}m{2cm}|>{\centering\arraybackslash}m{2cm}|}
     \hline
     n qubit & Rigetti Ankaa-3 with 4096 shots & Rigetti Ankaa-3 with 8192 shots  \\
     \hline
     2 & $\sim$~$0.19$ & $\sim$~$0.19$ \\
     \hline
     3 & $\sim$~$0.52$ & $\sim$~$0.51$ \\
     \hline
     4 & $\sim$~$0.55$ & $\sim$~$0.6$ \\
     \hline
     5 & $\sim$~$0.77$ & $\sim$~$0.75$\\
     \hline
    \end{tabular}
    
    \label{tab:RealHW2}
\end{table}

\section{Conclusions}
\label{lab:conclusion}
In this work, a novel QNN is proposed to address the state preparation problem in its relaxed version, with a scalable VQC designed on the diagonal $\mathfrak{su}(2^n)$ Lie algebra. The parameterized quantum circuit is derived from the traditional QSP ladder structure, redesigned through the Standard Recursive Block Basis (SRBB) decomposition to link the variational parameters to the topology of the Lie group $SU(2^n)$, thereby improving its expressiveness and approximation capabilities. By choosing the uniformly controlled gates (UCGs) accordingly, it is possible to use only the diagonal elements of SRBB, reducing the depth by an exponential factor compared to the full algebraic basis and achieving competitive depths and gate counts in light of algorithmic simplicity and scalability.

The algorithm has been implemented using the PennyLane library, and tests have been performed both with a simulator and with real quantum computers, specifically IQM Garnet, IQM Deneb, Rigetti Ankaa-3, and IonQ Forte1. States of different sizes have been considered, from 2 to 8 qubits with the simulator and from 2 to 5 qubits with the real quantum computers. 

The chosen algebraic decomposition leads to the preparation of the real part of the desired state separately from the respective complex part, so that the QNN is composed of two components: the first component approximates the modulus of each entry of the target state, while the second component approximates the corresponding phases. Training is then split into two steps; otherwise, the task becomes more complex, and the network cannot achieve good results. Two different optimizers, Adam and Nelder Mead, are considered. With the Adam optimizer, the trace distance and the fidelity are used as loss functions; however, when the Nelder-Mead optimizer is chosen, the Frobenius norm is minimized. 

Both random states and specific states have been tested. In simulation, the network achieves high accuracy (with errors ranging from $10^{-13}$ to $10^{-15}$) relative to the target state for up to 4 qubits. Accuracy starts to decrease with a higher number of qubits, though the error in the probabilities is within the third or fourth decimal place. 

When using real hardware, the QNN achieves good results with a small number of qubits when random states are chosen. However, the error is not fixed within a range; instead, it varies significantly depending on the target state. Furthermore, when the target state is the Bell state or a state with a uniform probability distribution, the network achieves a Hellinger distance of around $0.1$ with respect to the target state. In contrast, when a different sparse state is used as the target, the QNN produces poor results. The same is observed with 3 qubits: with GHZ and a state with a uniform probability distribution, the QNN achieves a distance of $0.38$ and $0.1$, respectively, while for other sparse states, the distance increases to $0.6$. Furthermore, on Rigetti Ankaa-3, the QNN is also tested with 4096 and 8192 shots to evaluate how the error varies, but no meaningful change in performance is observed.

In the future, the proposed framework could be compared with well-known QSP algorithms and quantum encoding techniques, such as Amplitude Encoding. Moreover, improving the SRBB synthesis scheme, along with optimizing the rotation gates, could further reduce the total depth. Another important task is to analyze the landscape, as the network can struggle to achieve high accuracy with a larger number of qubits due to the increased complexity of the parametric surface, which could lead to the emergence of the Barren Plateaus~\cite{mcclean2018barren} problem.



\subsection*{Code Availability}
The code is available at \url{https://github.com/qslab-unipr/aquaman-sp}

\begin{acks}
Giacomo Belli acknowledges financial support from the European Union - NextGenerationEU, PNRR MUR project PE0000023-NQSTI. 
Michele Amoretti acknowledges the financial support from Spoke 10 - ICSC - ``National Research Centre in High Performance Computing, Big Data and Quantum Computing'', funded by European Union – NextGenerationEU. This research benefits from the High Performance Computing facility of the University of Parma, Italy (HPC.unipr.it).
\end{acks}

\bibliographystyle{ACM-Reference-Format}
\bibliography{references}


\end{document}